\documentclass[a4paper,fleqn,usenatbib]{mnras}

\usepackage{newtxtext,newtxmath}

\usepackage[T1]{fontenc}
\usepackage{ae,aecompl}

\usepackage{graphicx}	
\usepackage{amsmath}	
\usepackage{amssymb}	

\newcommand{\OmegaBar}{\Omega_{\mathrm{b}}}
\newcommand{\diff}{\mathrm{d}}

\title[The Hercules stream explained by resonances]
{Trimodal structure of Hercules stream explained by originating from bar resonances}

\author[T. Asano et al.]{
Tetsuro Asano,$^{1}$\thanks{E-mail: t.asano@astron.s.u-tokyo.ac.jp}
M. S. Fujii,$^{1}$
J. Baba,$^{2}$
J. B{\' e}dorf,$^{3,4}$
E. Sellentin$^{3}$, and 
S. Portegies Zwart$^{3}$
\\
$^{1}$Department of Astronomy, Graduate School of Science, 
    The University of Tokyo, 7-3-1 Hongo, Bunkyo-ku, Tokyo, 113-0033, Japan\\
$^{2}$National Astronomical Observatory of Japan, Mitaka-shi, Tokyo 181-8588, Japan\\
$^{3}$Leiden Observatory, Leiden University, NL-2300RA Leiden, The Netherlands\\
$^{4}$Minds.ai, Inc., Santa Cruz, the United States
}

\date{Accepted XXX. Received YYY; in original form ZZZ}

\pubyear{2020}

\begin{document}
\label{firstpage}
\pagerange{\pageref{firstpage}--\pageref{lastpage}}
\maketitle

\begin{abstract}
{\it Gaia} Data Release 2 revealed detailed structures of nearby stars in phase space. These include the Hercules stream, whose origin is still debated.
Most of the previous numerical studies conjectured that the observed structures originate from orbits in resonance with the bar, based on static potential models for the Milky Way. We, in contrast, approach the problem via a self-consistent, dynamic, and morphologically well-resolved model, namely a full $N$-body simulation of the Milky Way. Our simulation comprises about 5.1 billion particles in the galactic stellar bulge, bar, disk, and dark-matter halo and is evolved to 10\,Gyr. Our model's disk component is composed of 200 million particles, and its simulation snapshots are stored every 10\,Myr, enabling us to resolve and classify resonant orbits of representative samples of stars. After choosing the Sun's position in the simulation, we compare the distribution of stars in its neighborhood with {\em Gaia}'s astrometric data, thereby establishing the role of identified resonantly trapped stars in the formation of Hercules-like structures. From our orbital spectral-analysis we identify multiple, especially higher order resonances. Our results suggest that the Hercules stream is dominated by the 4:1 and 5:1 outer Lindblad and corotation resonances. In total, this yields a trimodal structure of the Hercules stream. From the relation between resonances and ridges in phase space, our model favored a slow pattern speed of the Milky-Way bar (40--45\,$\mathrm{km \; s^{-1} \; kpc^{-1}}$).
\end{abstract}




\section{Introduction}
The European Space Agency (ESA) is operating an astrometric mission {\it Gaia} \citep{GaiaCollaboration2016} which observed our Milky Way.
Its second data release \citep[{\it Gaia} DR2;][]{GaiaCollaboration2018} provides five astrometric parameters for 1.3 billion sources and additional line of sight velocities for 7.2 million sources.
For stars, these astrometric data provide snapshots of their orbits in the Galactic potential and we here aim to obtain information on the dynamical structure of the Milky Way (MW) by investigating the phase-space distributions of its stars.
In this context, velocity-space structures of the solar neighborhood have previously been repeatedly studied \citep[e.g.][]{Kalnajs1991,Dehnen1998,Dehnen1999b}, but mainly under analytical or numerical approximations which we here drop. 

\begin{figure}
    \centering
    \includegraphics[width=\columnwidth]{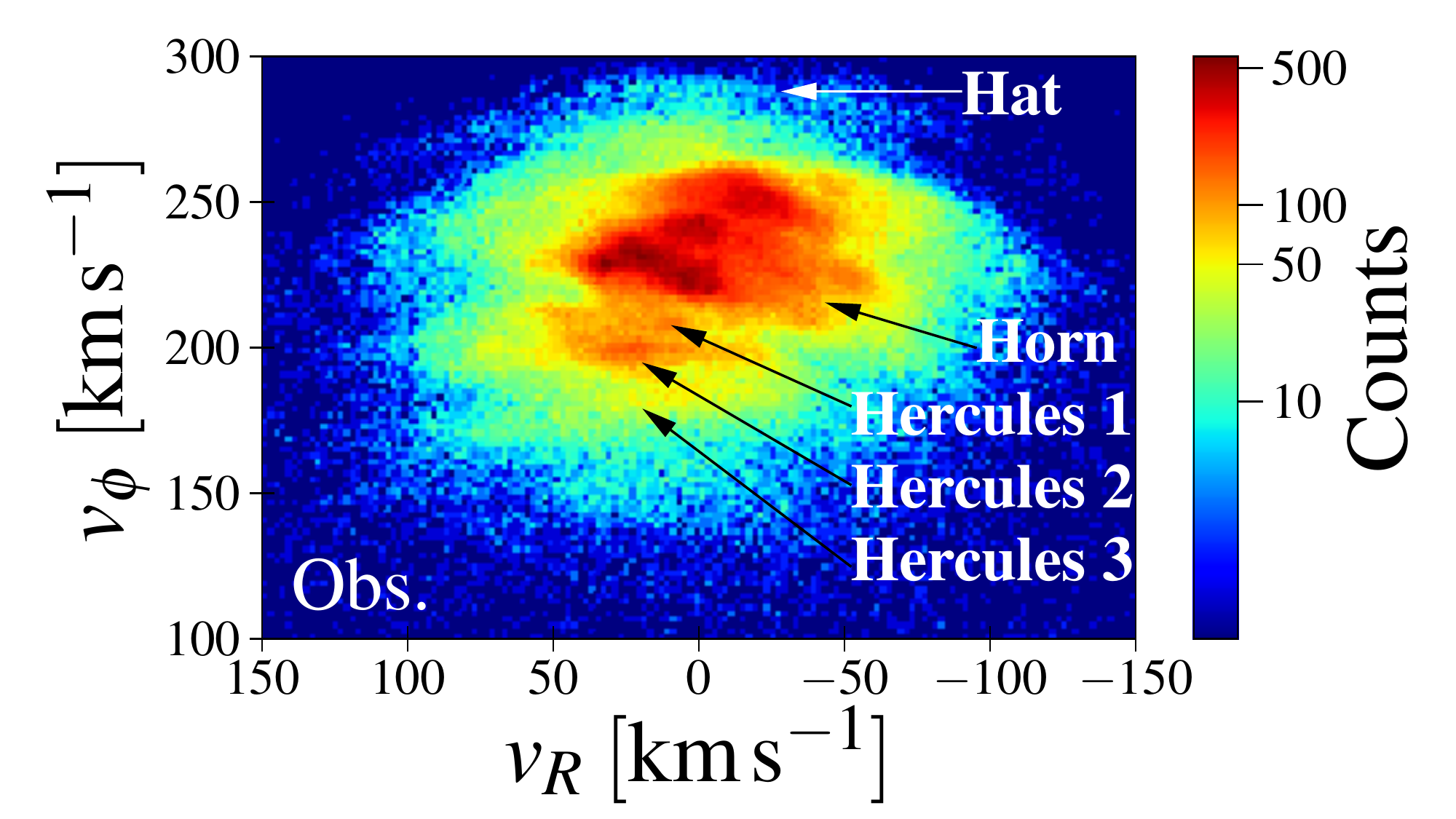}
    \caption{Radial and angular velocity-space distribution of stars within 0.2~kpc from the Sun, binned by $2\;\mathrm{km \; s^{-1}} \times 2\;\mathrm{km \; s^{-1}}$. From {\it Gaia} DR2 catalogue we selected stars whose relative errors in parallax are smaller than 10\%. 
    }
		\label{fig:uv_gaia}
\end{figure}

In Fig.~\ref{fig:uv_gaia}, we present the distribution of radial velocities, $v_R$, versus angular velocities, $v_{\phi}$, of stars observed with {\it Gaia} \citep[similar to Figure 22 in][]{Gaia-Katz+2018}. In this map, we see several arch-shaped over-densities with such as the so-called `horn' near  $(v_R,v_{\phi}) \sim (-50, 220) \; \mathrm{km\;s^{-1}}$ \citep{Monari+2017a,Fragkoudi+2019} and the so-called `hat' near $(v_R,v_{\phi}) \simeq (\pm 50, >250) \; \mathrm{km\;s^{-1}}$ \citep[e.g.][]{HuntBovy2018,Hunt+2019}. The most prominent structure is called the `Hercules' stream, and is located between $(v_R,v_{\phi}) \simeq (100, 200) \; \mathrm{km\;s^{-1}}$ and  $(v_R,v_{\phi}) \simeq (-70, 170) \; \mathrm{km\;s^{-1}}$.

\citet{Dehnen1998} originally identified the Hercules stream as $U$-anomaly from the {\it Hipparcos} data \citep{ESA1997, Perryman+1997}, and {\it Gaia} DR2 \citep{GaiaCollaboration2018, Gaia-Katz+2018} revealed further detailed structures within. 
{\it Gaia} DR2 showed for the first time the trimodal structure of the Hercules stream, which is composed of three sub-streams at $v_{\phi} \simeq 220$, 200, and $180\,\mathrm{km\;s^{-1}}$ \citep{Gaia-Katz+2018,Ramos+2018,Trick+2019,LiShen2020}. 
Although we refer to the bottom stream in Fig.~\ref{fig:uv_gaia} as `Hercules 3', it is identical to a moving group, HR 1614 \citep{Eggen1971, Eggen1978, Kushniruk+2020}.

Soon after the first discovery of the Hercules stream, a scenario on its origin was proposed by \citet{Dehnen1999b,Dehnen2000}.
He calculated the evolution of the stellar phase space distribution in the simple $m=2$ bar potential (where $m$ indicates the azimuthal Fourier mode), and then argued that stars trapped in the 2:1 outer Lindblad resonance (OLR) of the bar can create moving groups in the solar neighborhood.
The exact position of the formed structure in phase space depends on the location of the OLR and the OLR position depends in turn on the pattern speed of the bar ($\OmegaBar$). Hence, to reproduce the observed features, this scenario requires a fast rotating bar to bring the 2:1 OLR around 8\;kpc, i.e. $\OmegaBar/\Omega_0 = 1.85$ (where $\Omega_0$ is the local circular frequency); this corresponds to $\OmegaBar= 53 \pm 3 \; \mathrm{km \; s^{-1} kpc^{-1}}$ \citep[see also][]{Fux2001,Minchev+2007,Minchev+2010,Antoja+2014}.
Such a fast ($\OmegaBar\sim 50\; \mathrm{km \; s^{-1} kpc^{-1}}$) bar model is also favored by test particle integration in a more realistic Milky Way (MW) Galaxy potential that was constructed from an $N$-body simulation \citep{Fragkoudi+2019}.

On the other hand, recent observations suggests slower pattern speeds of the bar.
Combining data from {\it Gaia} DR2 and further surveys, both \citet{Sanders+2019} and \cite{Bovy+2019} estimated a pattern speed of $\OmegaBar = 41 \pm 3 \; \mathrm{km \; s^{-1} \; kpc^{-1}}$, and \citet{Clarke+2019} estimated $\OmegaBar = 37.5\; \mathrm{km \; s^{-1} \; kpc^{-1}}$. 
Recent measurements of the bar length also support a slow bar model. The measured ratio of the corotaion radius ($R_{\rm CR}$) to the bar length ($R_{\rm b}$) $R_{\mathrm{CR}}/R_{\mathrm{b}} = 1.2 \pm 0.2$ for external galaxies \citep{Aguerri+2003, Aguerri+2015}. The bar length of $R_{\mathrm{b}} = 5.0 \pm 0.2$~kpc in the Milky Way \citep{Wegg+2015}, therefore, suggests the corotation radius of $R_{\mathrm{CR}} \simeq$ 5--7~kpc or the pattern speed of $\OmegaBar \simeq$ 34--47~$\mathrm{km \; s^{-1} \; kpc^{-1}}$. 
In addition, such slow and long bar is also supported by the dynamical modelling of bulge stars \citep{Portail+2017} and gas kinematics \citep{Sormani+2015c,Li+2016}.
With such a slow pattern speed, the 2:1 OLR should be located further than 8\,kpc.

The slow and long bar model enhanced the studies of the Hercules stream's origin, and new scenarios have been suggested. Most of them agree with the concept that resonant orbits due to non-axisymmetric structures, such as a bar and/or spiral arms, create the moving groups in the solar neighborhood revealed by {\it Gaia} DR2. 
One of the new scenarios is the bar's corotation (CR) scenario. \citet{Perez-Villegas+2017} integrated orbits of test particles in a gravitational field constructed from a self-consistent $N$-body model of the Milky Way to match observational data using the Made-to-Measure (M2M) method \citep[][]{Portail+2017}, and then proposed that the Hercules-like stream can be made of stars orbiting around Lagrangian points of the bar. These stars move outward from the bar's CR radius to visit the solar neighbourhood. In this study, they assumed a slow bar ($\OmegaBar \sim 40 \; \mathrm{km \; s^{-1} \; kpc^{-1}}$). This scenario is also supported by analytic calculations of perturbed distribution functions in resonance regions \citep{Monari+2019a,Monari+2019b,Binney2020}, and full $N$-body simulations of a MW-like galaxy \citep{D'OnghiaAguerri2019}.

Another new scenario attributes the Hercules stream to higher-order resonances of the slow, long bar.
\citet{HuntBovy2018} integrated orbits of test particles in analytic bar potentials including an $m=4$ Fourier mode and suggested that the 4:1 OLR of a slowly rotating bar can lead to a bi-modal structure in the Hercules stream. \citet{Monari+2019a} also studied the relation between the bar's higher-order resonances and the solar neighborhood kinematics using a resonant distribution function model \citep{Monari+2017b} in the same realistic bar potential as that used in \citet{Perez-Villegas+2017}. They showed that CR and 6:1 OLR of the bar create a Hercules-like stream and a horn-like structure, respectively.
Similarly, \citet{Michtchenko+2018a} performed test particle simulations in a slow-bar potential with higher-order multipole moments combined with a spiral potential, and suggested that the Hercules stream associated with the spiral's 8:1 ILR \citep[see also][]{Barros+2020}. On the other hand, \citet{Hattori+2019} also showed that Hercules-like streams originate from orbit families associated with the 5:1 inner Lindblad resonance (ILR) in the simple $m=2$ slow-bar + spiral potentials. 

Note that most of the above studies are based on test particle simulations in `static' (or analytic) potentials.  
We can easily capture resonant orbits of stars in static potentials; however, previously performed self-consistent $N$-body simulations of (barred) spiral galaxies have suggested that the structures in the spiral arms and bar change with time in complicated ways \citep[e.g.][]{1984ApJ...282...61S,SellwoodSparke1988,Baba+2009,Baba+2013,Grand+2012a,Grand+2012b,D'Onghia+2013,Fujii+2011,Khoperskov+2019}. Such transient nature may affect stellar orbits in higher-order resonances. Higher order resonances are usually weaker than CR or 2:1 OLR. In order to detect higher order resonances such as 4:1 and 6:1 OLR in self-consistent $N$-body simulations, a high enough resolution is required.
Recently, \citet{D'OnghiaAguerri2019} captured stars in the CR resonance using fully self-consistent $N$-body simulations, but they did not report higher order resonances as the origins of the Hercules stream.
Hence, so far no high-order resonances have been reported in self-consistent $N$-body simulations.

Phase space structures have been found in these self-consistent $N$-body models, but their origin is unknown.
\citet{Fujii+2019} performed $N$-body simulations of disk galaxy models and found a MW-like model, of which some observed structures and kinematics match those of the MW. They performed simulations using a maximum of eight billion particles (more than two hundred million particles for the disk) with the dark-matter halo modeled with $N$-body particles (live halo). This is an order of magnitude higher mass-resolution than previous similar studies \citep[e.g.][]{D'OnghiaAguerri2019}. \citet{Fujii+2019} showed $U$-$V$ maps obtained in their simulations and discussed if Hercules-like streams are found in $N$-body simulations. Indeed, they found some Hercules stream-like structures in their simulations, but the origin of the structures has not been investigated.

In this paper, we analyze the results of the $N$-body simulations performed in \citet{Fujii+2019}, which represents an isolated Milky Way-like galaxy modeled completely as an $N$-body system. In this simulation, we track and classify the stellar orbits. 
The purpose of the classification is to identify stars trapped in resonance. 
We then show that stars trapped in higher-order resonances actually exist in the live disk potential and that the trimodal structure of the Hercules stream can be explained by such higher-order resonances.


\section{Milky Way $N$-body simulations}
\subsection{Initial condition and dynamical evolution of the Milky-Way model}
We use one of the Milky Way $N$-body simulations that were performed by \citet{Fujii+2019}. Here, we describe their model and simulation methods.

We use model MWa of \citet{Fujii+2019}. This model is a Milky Way-like galaxy composed of a live stellar disk, a live classical bulge, and a live dark-matter (DM) halo, and the initial conditions were generated using {\tt GalactICS} \citep{KuijkenDubinski1995,WidrowDubinski2005}. The stellar disk follows an exponential profile with a mass of $3.73 \times 10^{10}M_{\sun}$, an initial scale-length ($R_{\rm d}$) of $2.3$~kpc, and an initial scale-height of $0.2$ pc. The classical bulge follows the Hernquist profile \citep{Hernquist1990}, whose mass and scale-length are $5.42 \times 10^9 M_{\sun}$ and $750$ pc, respectively. The DM halo follows the Navarro–Frenk–White (NFW) profile \citep{Navarro+1997}, whose mass and scale radius are $8.68 \times 10^{11}M_{\sun}$ and 10\,kpc, respectively. A more detailed model description can be found in \citet{Fujii+2019}.

The simulations were performed using the parallel GPU tree-code, {\tt BONSAI}\footnote{{\tt https://github.com/treecode/Bonsai}} \citep{Bedorf+2012,Bedorf:2014:PGT:2683593.2683600} with a GPU cluster, Piz Daint.

The simulation was started from a disk without any structures. After $\sim 2$\,Gyr, a bar started to form, and continued to grow until $\sim 5$\,Gyr. During the evolution, the bar slowed down with oscillations up to $\sim 8$\,Gyr. Spiral structures also formed, and are most prominent at $\sim 5$\,Gyr. However, they faint at later times due to the dynamical heating of the disk. The simulation was continued up to 10\,Gyr.

This simulation is one of the most highly resolved $N$-body simulations that we have performed. The numbers of particles of this models are {30M, 208M}, and  4.9B for the bulge, disk, and halo, and the same mass resolution is used for all the three components. This large number of particles enables us to perform a direct comparison of simulation results with observed data. A further advantage of this simulation is the high time resolution of the particle data output. The position- and velocity-snapshots of disk particles were stored every 9.76\,Myr. We therefore can directly recover the actual orbits of individual stars from the snapshots. 

\subsection{Determination of the bar's pattern speed in the simulation}

In this simulation we determine the bar's pattern speed using the Fourier decomposition as was also done in \citet{Fujii+2019}.
We divide the galactic disk into annuli with a width of 1~kpc, and then Fourier decompose the disk's surface density in each annulus:
\begin{equation}
	\Sigma (R, \phi) = 
	\sum_{m=0}^{\infty} A_m(R) \exp\{im[\phi - \phi_m(R)]\},
	\label{eq:Fourier_Decomp}
\end{equation}
where $A_m(R)$ and $\phi_m(R)$ are the $m$-th mode's amplitude and phase angle, respectively.  
We define $\phi_2(R)$ averaged in $R<3$~kpc as the angle of the bar in the snapshot \citep[see also][]{Fujii+2019}. We obtained the angle of the bar of the last 64 snapshots, which corresponds to $t=9.38$--$10$~Gyr.
Finally, the bar's pattern speed, $\OmegaBar$, is determined using the least squares fitting to $\phi_2 (t) = \OmegaBar t + \phi_{2, 0}$, where $\phi_{2, 0}$ is the angle of the bar in the first snapshot we used. 
The thus obtained pattern speed of the bar in the simulation is $\OmegaBar=46.12\; \mathrm{km\;s^{-1}\;kpc^{-1}} = 1.53 \Omega_{\mathrm{8kpc}}$ where $\Omega_{\mathrm{8kpc}}$ is the circular frequency at $R=8$~kpc in our MW model. 
Recent studies such as \citet{Sanders+2019} 
and \citet{Bovy+2019} suggested a pattern speed of the Galactic bar of $\OmegaBar = 41\; \mathrm{km\;s^{-1}\;kpc^{-1}} \simeq 1.4\Omega_0$, where $\Omega_0$ is the circular velocity at the solar distance. 
The bar's pattern speed in our simulation is slightly higher than this value, but slower than that favored by \citet{Dehnen2000}'s 2:1 OLR model for the Hercules stream ($\OmegaBar \simeq 50\; \mathrm{km\;s^{-1}\;kpc^{-1}} \simeq 1.8\Omega_0$).

\subsection{Distribution of simulated stars in velocity space}

\begin{figure*}
	\centering
	\includegraphics[width=\linewidth]{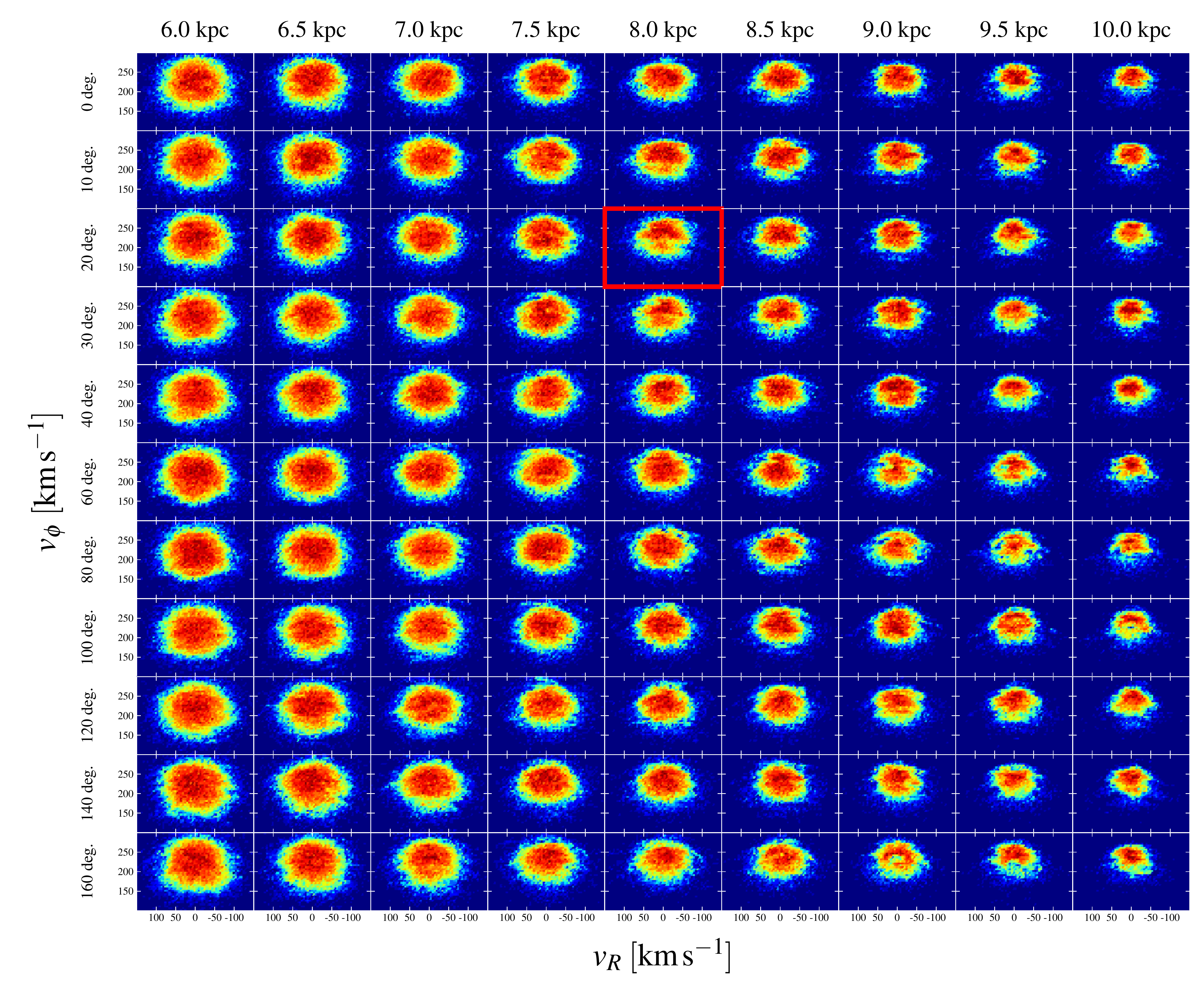}
	\caption{2D-histogram in $v_R$ versus $v_{\phi}$ space at various positions in the disk at 10\,Gyr. The distribution is for particles within $0.2$~kpc from each position. In all  panels, the bin size is set to $5\;\mathrm{km\;s^{-1}} \times 5\;\mathrm{km\;s^{-1}}$. The panel for $(R,\phi)=(8\,\mathrm{kpc},20^{\circ})$, corresponding to the solar neighborhood, is framed by a red rectangle.}
	\label{fig:uv_all}
\end{figure*}

\begin{figure*}
	\centering
	\includegraphics[width=\columnwidth]{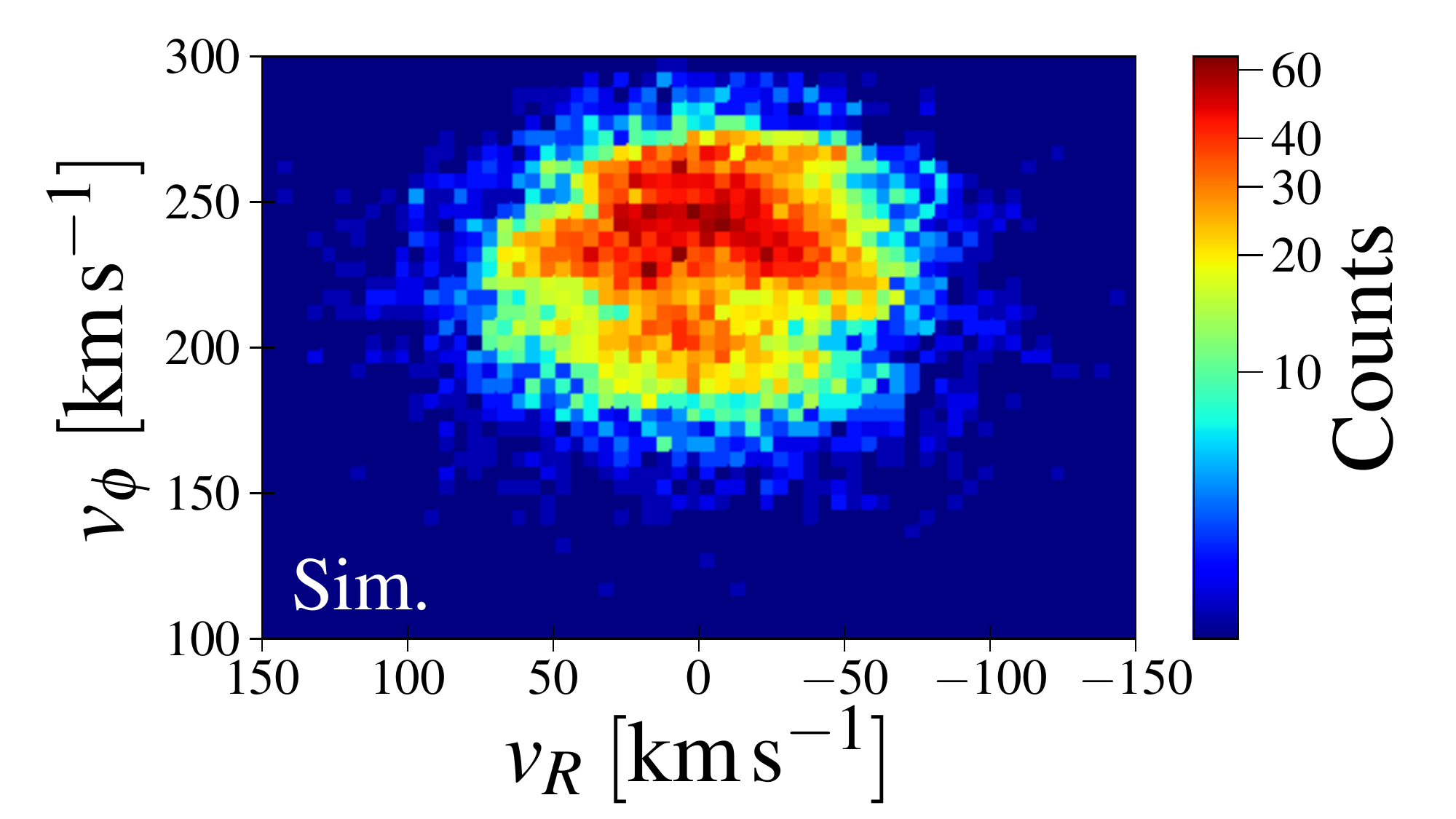}
	\includegraphics[width=\columnwidth]{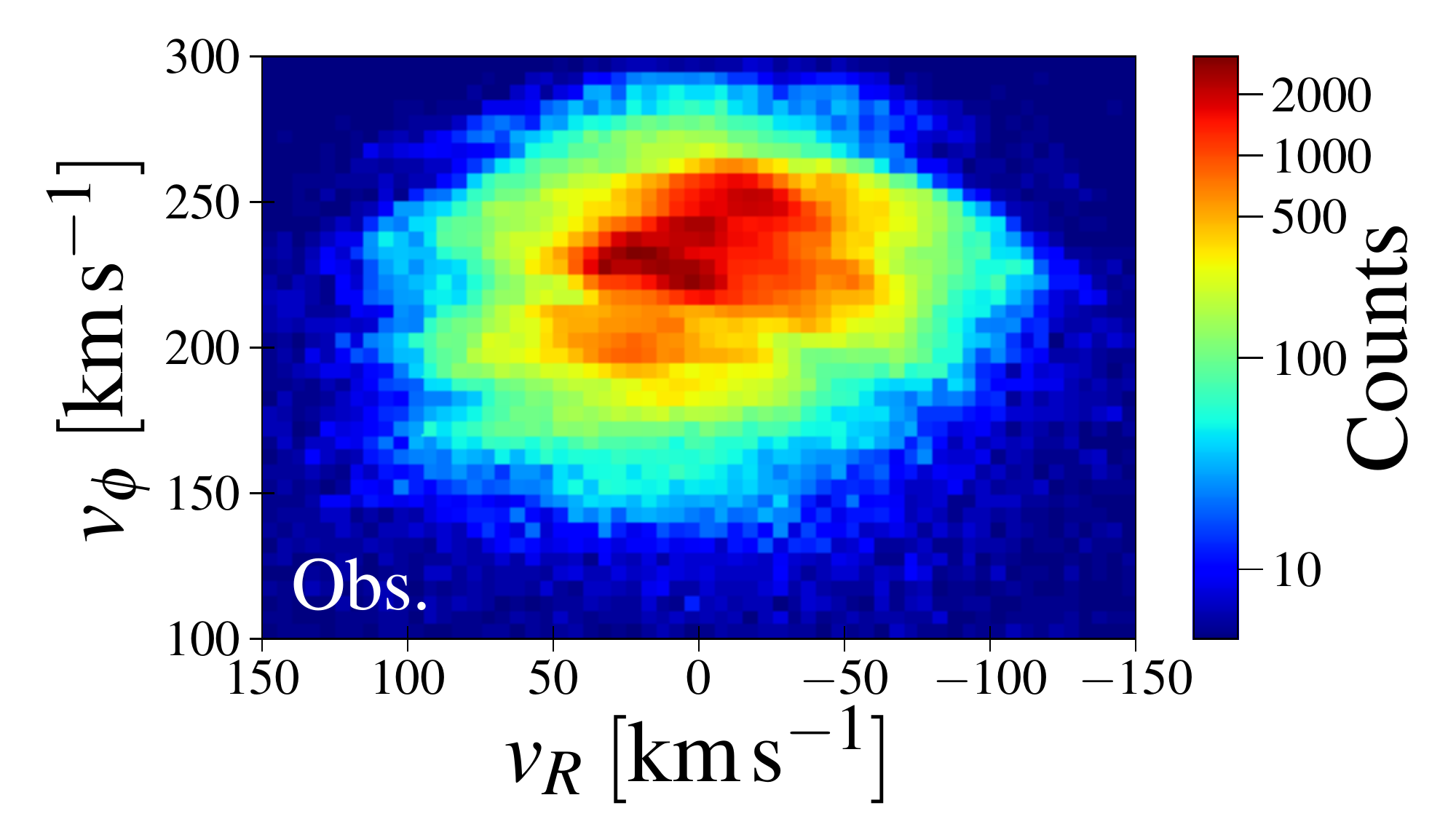}
	\caption{{\it Left:\/} Distribution of simulation particles in velocity space around $(R,\phi)=(8\,\mathrm{kpc},20^{\circ})$. This is an enlargement of the panel framed by a red rectangle in Fig.~\ref{fig:uv_all}. {\it Right:\/} The same figure from {\it Gaia} DR2. The total number of the stars in the maps of the simulation and the observation are 13,303 and 345,152, respectively.}
	\label{fig:uv_8_20}
\end{figure*}

To study the origin of the Hercules stream, we begin by locating a Sun-like position in the simulated galactic disk. We take the last snapshot of the simulations ($t=10$~Gyr) and plot the velocity-space distribution of particles within 0.2~kpc from the ``Sun'' and iterate over positions in the disk. The results are shown in Fig.~\ref{fig:uv_all}.  
Note that we assume the position of the ``Sun'' in the galactic mid-plane ($z=0$) in the MW model.

At certain locations, we find a velocity-space structure similar to the one observed in Gaia DR2 (\citealt{Gaia-Katz+2018} and Fig. \ref{fig:uv_gaia}). These observational studies suggest that the Sun has a distance of $R_{\sun}=(8.178 \pm 13 \pm 22)$~kpc \citep{GravityCollaboratio+2019} from the Galactic center and lies at an angle of $\phi_{\sun}=27^{\circ} \pm 2^{\circ}$ \citep{Wegg+2013}, $ 29^{\circ}\pm2^{\circ}$ \citep{Cao+2013}, $ 24^{\circ}$--$27^{\circ}$ \citep{Rattenbury+2007}, and $ 20^{\circ}$--$25^{\circ}$ \citep{Bissantz-Gerhard2002} with respect to the major axis of the Galactic bar. By analysing our simulations, we select a distance and  position relative to the bar of $(R, \phi)=(8\, \mathrm{kpc},20^{\circ})$.
This choice is based on the similarity of the structure observed in Fig.\,\ref{fig:uv_gaia} and compared with the kaleidoscope of similar phase-space images in Fig.\ref{fig:uv_all}.
 The velocity-space map at $(R, \phi)=(8\, \mathrm{kpc},30^{\circ})$ also shows the Hercules-like stream, but it is most clearly identified in the map at $(R, \phi)=(8\, \mathrm{kpc},20^{\circ})$. 
The space-coordinates framed in red in Fig.\ref{fig:uv_all} has the closest correspondence with the observed image.
We then select this particular phase-space position as the one to represent the Sun in our simulations.

Fig.~\ref{fig:uv_8_20} is the enlarged figure of the velocity-space distribution at $(R, \phi)=(8\, \mathrm{kpc},20^{\circ})$ (same as the panel framed by a red rectangle in Fig.~\ref{fig:uv_all}).
A Hercules-like stream is located from $(v_R, v_{\phi} )\simeq (80, 200 )\, \mathrm{km\; s^{-1}}$ to $(v_R, v_{\phi} )\simeq (-60, 190 )\, \mathrm{km\; s^{-1}}$ in this figure.
In addition, a Hat-like stream and a Horn-like structure are also seen from $(v_R, v_{\phi} ) \simeq (50, 260 )$ to $(v_R, v_{\phi} ) \simeq(-50, 270 )\, \mathrm{km\; s^{-1}}$ and around $(v_R, v_{\phi} ) = (-50, 220 )\, \mathrm{km\; s^{-1}}$, respectively.
Some other moving groups known from observations, such as the Hyades and Pleiades,  cannot be clearly mapped to structures in the simulated MW model.

\section{Orbit analysis and identification of resonantly trapped stars}\label{sec:orbit_analysis}
The key advantage of working with an $N$-body simulation is that resonances can be tested in live systems. The spiral structures and bars evolve with time in a live potential, which is not the case in static potential simulations. In such a time-dependent potential, stars in a resonance may not be able to stay stable in the resonance. In order to capture resonances in a live potential, the orbits of  individual stellar orbits have to be followed in the live simulations.

To this aim, we thus use the snapshots of our $N$-body simulation, and trace the orbits of the stars that we are interested in. We especially determine the orbital frequencies of the disk particles in order to classify the particles by their orbital characteristics such as resonances. 
For this purpose, we perform a frequency analysis of stellar orbits obtained from the $N$-body simulation with a following frequency measurement method used in \citet{2007MNRAS.379.1155C}.
Here, we detail upon the classification scheme used to identify stellar properties in our simulation. 

First, we determine the radial frequency, $\Omega_R$, using the Discrete Fourier Transformation (DFT) for $R(i)$ where $R(i) \; (i=1,\ldots,64)$ is a radial coordinate in the $i$-th snapshot. 
We employ a zero-padding technique for Fourier transforming:\  960 zero points are added at the end of the data series.
We then sample frequency space with 512 points between $0\; \mathrm{km\; s^{-1}\; kpc^{-1}}$ and
$315\; \mathrm{km\; s^{-1}\; kpc^{-1}}$, whereby the upper bound is given by the Nyquist frequency.
We identify a resonant $\Omega_R$ as a frequency that causes a local maximum in the Fourier spectrum: obviously the local maximum indicates an over-abundance of stars at this frequency.
In contrast, the associated angular frequency $\Omega_{\phi}$ is determined by a regression analysis instead of a Discrete Fourier Transform.
From the snapshots, we collect per particle the series of measured angles $\phi(i)$ as a function of time $t(i)$, where $i$ iterates over the 64 snapshots available. For each particle, this results in pairs $[t(i), \phi(i)]\; (i=1,\ldots,64)$ to which we fit the function $\phi = \Omega_{\phi} t +\phi_0$ using a least squares method. The resulting Least-Squares Estimator yields the angular frequency $\Omega_{\phi}$ of the studied star.

\begin{figure}
	\centering
	\includegraphics[width=\columnwidth]{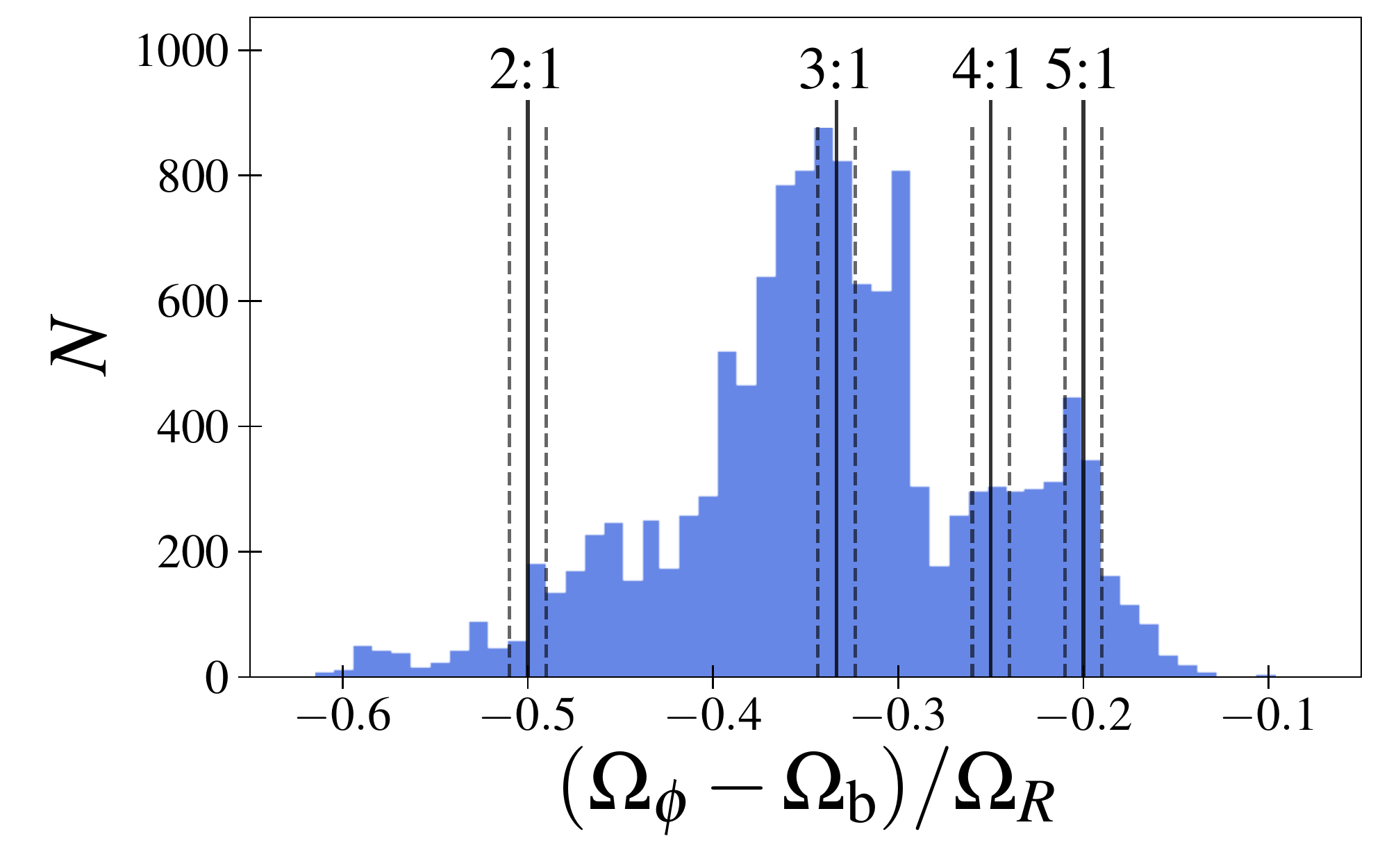}
	
	\caption{Orbital frequency ratios for the particles within 0.2~kpc from $(R, \phi)=(8\, \mathrm{kpc}, 20^{\circ})$. The four vertical solid lines indicate $(\Omega_{\phi} - \OmegaBar)/\Omega_R =$ -1/2, -1/3 -1/4, and -1/5, corresponding to the 2:1, 3:1, 4:1, and 5:1 resonances, respectively.  Pairs of dashed lines beside solid lines indicate frequency ratios of $-1/2 \pm 0.01$, $-1/3 \pm 0.01$, $-1/4 \pm 0.01$, and $-1/5 \pm 0.01$, respectively. For each resonance, particles whose frequency ratios are between a dashed line pair are selected as resonantly trapped particles. }
	\label{fig:freq_ratio}
\end{figure}

In Fig.~\ref{fig:freq_ratio},
we show the distribution of the frequency ratio $(\Omega_{\phi} - \OmegaBar)/\Omega_R$ for the particles within 0.2~kpc from our ``Sun's'' location at $(R, \phi)=(8\, \mathrm{kpc}, 20^{\circ})$. 
Here, we use the bar's pattern speed ($\OmegaBar$) obtained from our simulations (see Section 2.2). 
The clearly visible statistically significant peaks in the distribution correspond to stars in resonances with the bar, whereas small fluctuations simply arise from binning. In Fig.~\ref{fig:freq_ratio}, we indicate the positions of multiple outer Lindblad resonances (OLR) as vertical solid lines. In the region that we assumed as the solar neighborhood, we find at the four OLR of 2:1, 3:1, 4:1, and 5:1.

We assume that particles within a rage of $\pm 0.01$ from the exact resonance frequency ratio are also in resonance; these are the two adjacent bins to the exact resonance frequency. We select these as resonant particles in the following analyzes. As an example, for particles trapped in the 2:1 resonance, we select those whose frequency ratios are in the range of $-0.51<(\Omega_{\phi} - \OmegaBar)/\Omega_R < -0.49$ which corresponds to the bins between the two vertical dashed lines next to the line of $(\Omega_{\phi} - \OmegaBar)/\Omega_R = -0.5$.

\begin{figure*}
	\centering
	\includegraphics[width=\linewidth]{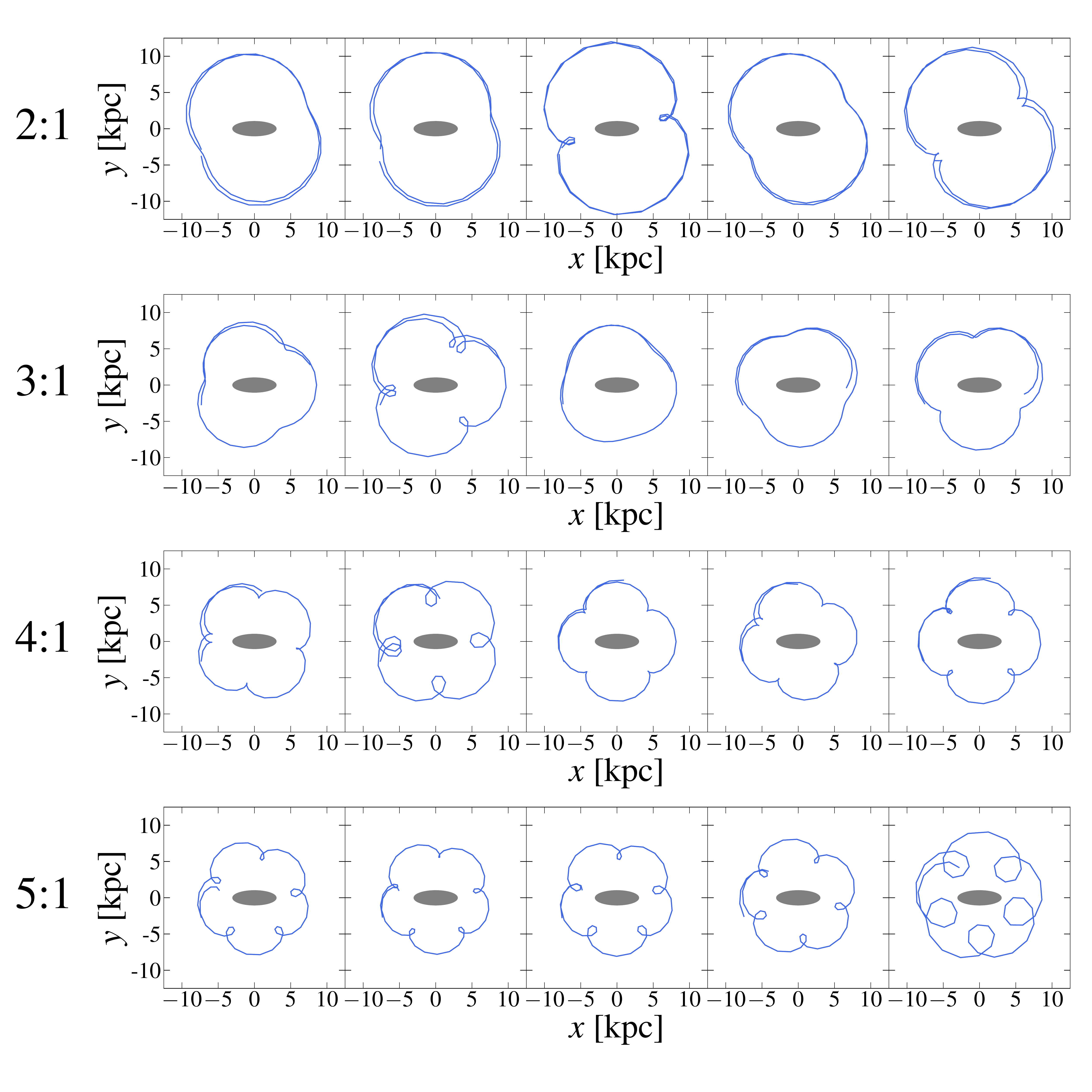}
	\caption{Examples of the orbits trapped in the bar resonances. The gray ellipse in the figure represents the bar orientation. The orbits are shown in the bar's rotating frame. The galaxy rotates clockwise, hence particles rotate counter-clockwise in this frame.}
	\label{fig:orbit_examples}
\end{figure*}

\begin{figure}
	\centering
	\includegraphics[width=\columnwidth]{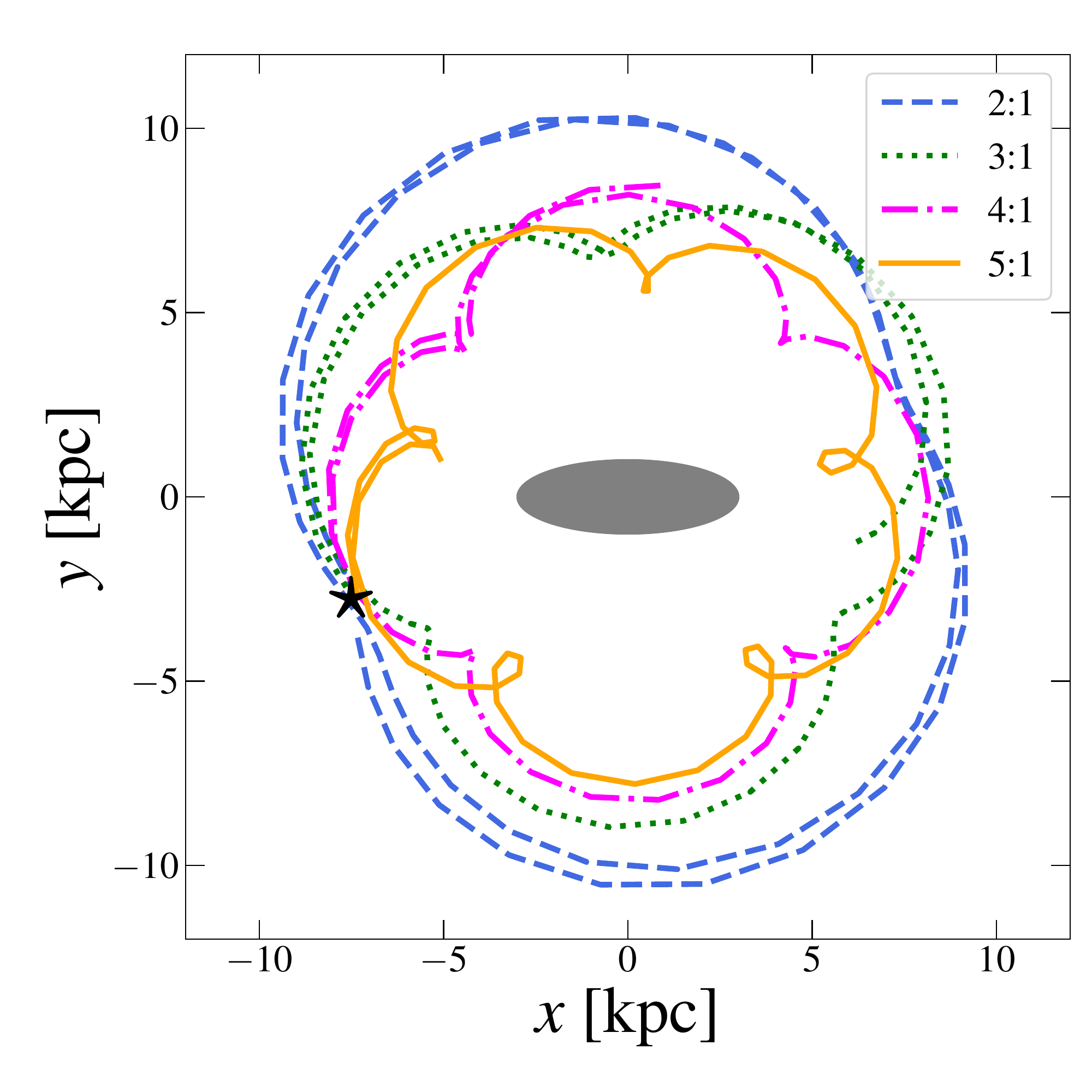}
	\caption{Resonantly trapped orbits in a single frame. The blue dashed, green dotted, magenta dash-dotted, and orange solid lines represent the 2:1, 3:1, 4:1, and 5:1 resonance orbits, respectively. In this figure, the galaxy rotates clockwise. The star symbol presents the position of the Sun in our model, $(R, \phi)=(8\, \mathrm{kpc},20^{\circ})$.}
	\label{fig:orbits_single_frame}
\end{figure}

In order to confirm the validity of this selection procedure, we verify whether the selected particles are  in resonant orbits. to verify the results of the Fourier analysis we randomly select 100 particles from the various resonant areas and plot their orbits over time. By visually inspecting them, as the sub-sample of 5 cases for each of the orbital resonances, presented in Fig.~\ref{fig:orbit_examples}, we confirm that except a few odd cases, all stars are on the appropriate resonance orbit. We show the panels that demonstrate this procedure for 100 different plotted orbits in the online material.
All the orbits in Fig.~\ref{fig:orbit_examples} are shown in the rotating frame of the bar, which is represented by the gray ellipse in each panel. In this figure, the galaxy rotates clockwise. In this frame, the particles accordingly rotate counter-clockwise as their angular frequencies are slower than the bar's pattern speed.
Stars trapped in the $m$:1 resonances oscillate $m$ times in the radial direction while they circle the bar. Thus, we confirm that our method properly identifies stars in resonant orbits. 

Note that naively, one would expect that orbits of the 2:1 resonance to align with the bar or to be perpendicular to it in the case of OLR.
In our simulation, however, we find that the orbits in the 2:1 OLR are inclined with respect to the orientation of the bar as shown in the top five panels in Fig.~\ref{fig:orbit_examples}. In other regions, however, particles which are exactly aligned with or perpendicular to the bar are dominant. In the region which we assumed as the solar neighborhood, particles in inclined orbits are dominant. 
The reason is currently not clear, but this may be because the position we "observed" is not exactly on the 2:1 OLR radius. Other possible reasons are spiral arms and/or the bar's slow-down.

Orbital radii decrease from lower to higher order resonances (from the top to the bottom in Fig.~\ref{fig:orbit_examples}).
This is more apparent in Fig.~\ref{fig:orbits_single_frame}, in which
we show four resonantly trapped orbits in a single frame. This is simply because higher-order resonances have smaller guiding radii.

\section{Results for Gaia DR2}
Using the orbit analysis of our simulation detailed upon in Sect.~\ref{sec:orbit_analysis}, we succeeded in identifying the simulation's bar pattern speed, and the resonance type of stars trapped in resonant orbits. By now comparing to the Gaia data, we can thus use our simulation analysis to constrain the properties of the Milky-Way galaxy. The thus found implications for the origin of the Hercules stream are presented in Sect.~\ref{Ssect:Hercules}, and constraints on the Milky Way's bar pattern speed are presented in \ref{Ssect:Bar}.

\subsection{The origin of the Hercules stream from resonances}
\label{Ssect:Hercules}
\begin{figure}
	\centering
	\includegraphics[width=\columnwidth]{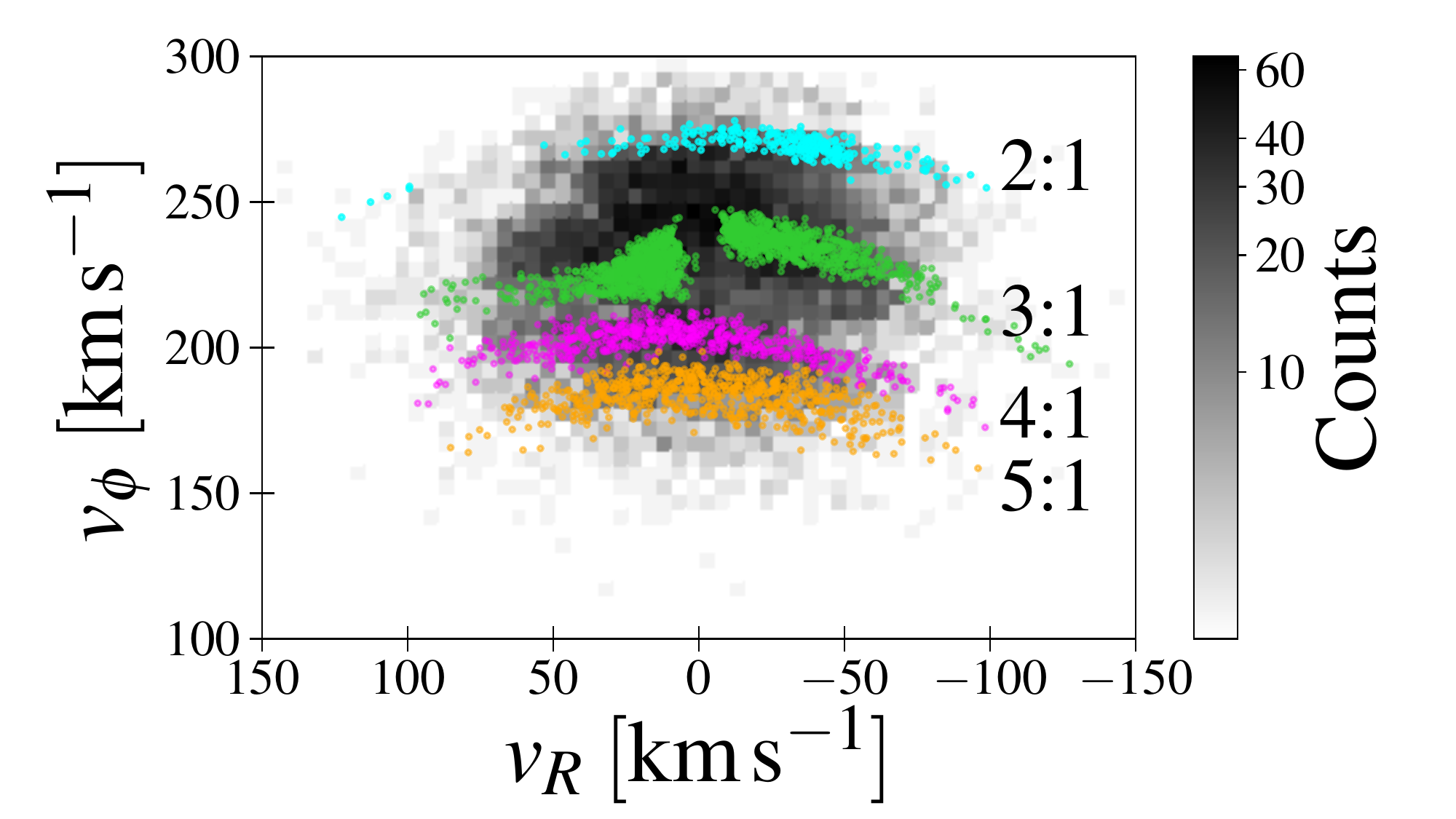}
	\includegraphics[width=\columnwidth]{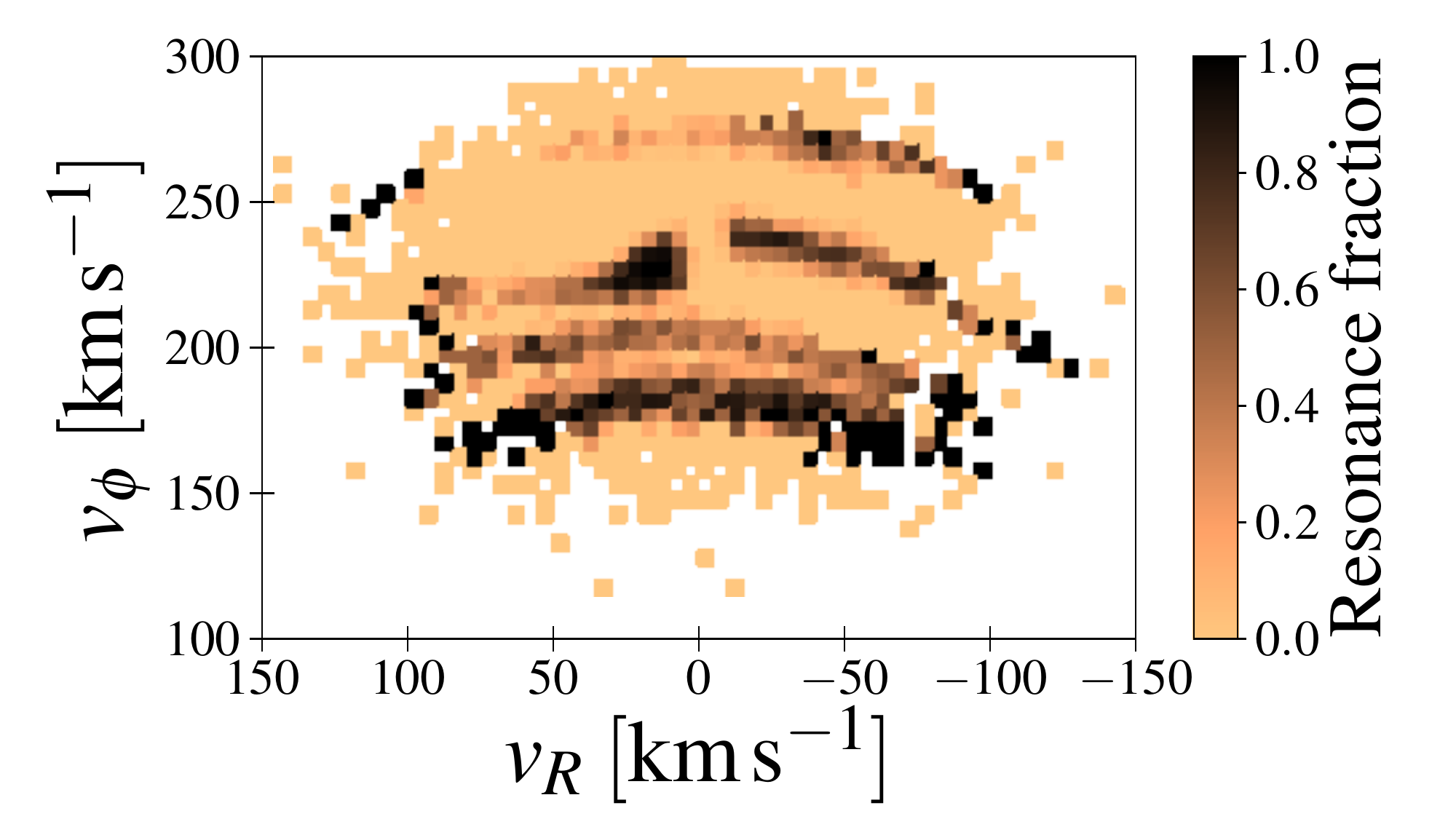}
	\caption{{\it Top:\/} Resonantly trapped particles in velocity space. The velocities of resonantly trapped particles are overplotted on the $v_R$ versus $v_{\phi}$ plane in Fig.~\ref{fig:uv_8_20}. Cyan, green, magenta, and orange dots represent particles trapped in the 2:1, 3:1, 4:1, and 5:1 resonances, respectively. {\it Bottom:\/} Colors indicate the fraction of the particles trapped in the 2:1, 3:1, 4:1, or 5:1 resonances in each bin.}
	\label{fig:uv_resonance}
\end{figure}

The \emph{Gaia} DR2, the Hercules stream creates a prominent overdensity of stars in the $v_R$ versus $v_{\phi}$ projection of the \emph{Gaia} data (see Fig.~\ref{fig:uv_gaia}). The recognition of this overdensity's trimodality \citep[e.g.][]{Ramos+2018} is recent, and requires an explanation. We here present the evidence that stars trapped in resonant orbits will form such a trimodal stream structure.

This can be seen from the top panel of Fig.~\ref{fig:uv_resonance}, where we over plot the position of particles in resonance in the velocity map (Fig.~\ref{fig:uv_8_20}). In the bottom panel of the figure, we show the velocity map colored by the fraction of particles in resonance in each bin. We find that between 60\% and 100\% of all stars in the Hercules-like stream of our simulation are trapped in the 4:1 and 5:1 OLR. This implies that the stars within the
Hercules stream (two streams at $v_{\phi}=220$ and 200\,km\,s$^{-1}$) will also be dominated by resonantly trapped stars.
Although the two resonantly trapped families are blurred in our velocity map of Fig.~\ref{fig:uv_8_20}, these two streams light up and are clearly separated when classified by resonance type in Fig.~\ref{fig:uv_resonance}, thereby providing a natural explanation for  multiple streams of \emph{Gaia}'s observations of the Hercules stream.

In our simulations, stars trapped in the 2:1 OLR have velocities much higher than the rotation speed at the position of the Sun. This is because the guiding center of the 2:1 OLR in our model is outer than 8\;kpc. These stars appear to populate the area in phase space that was characterized by the `hat' structure in the observation. Stars in the 3:1 OLR are distributed around the circular velocity at 8\;kpc. Among the 3:1 OLR stars, those with $v_{\rm R}<0$ are part of the  
`horn' structure.

\begin{figure}
	\centering
	\includegraphics[width=\columnwidth]{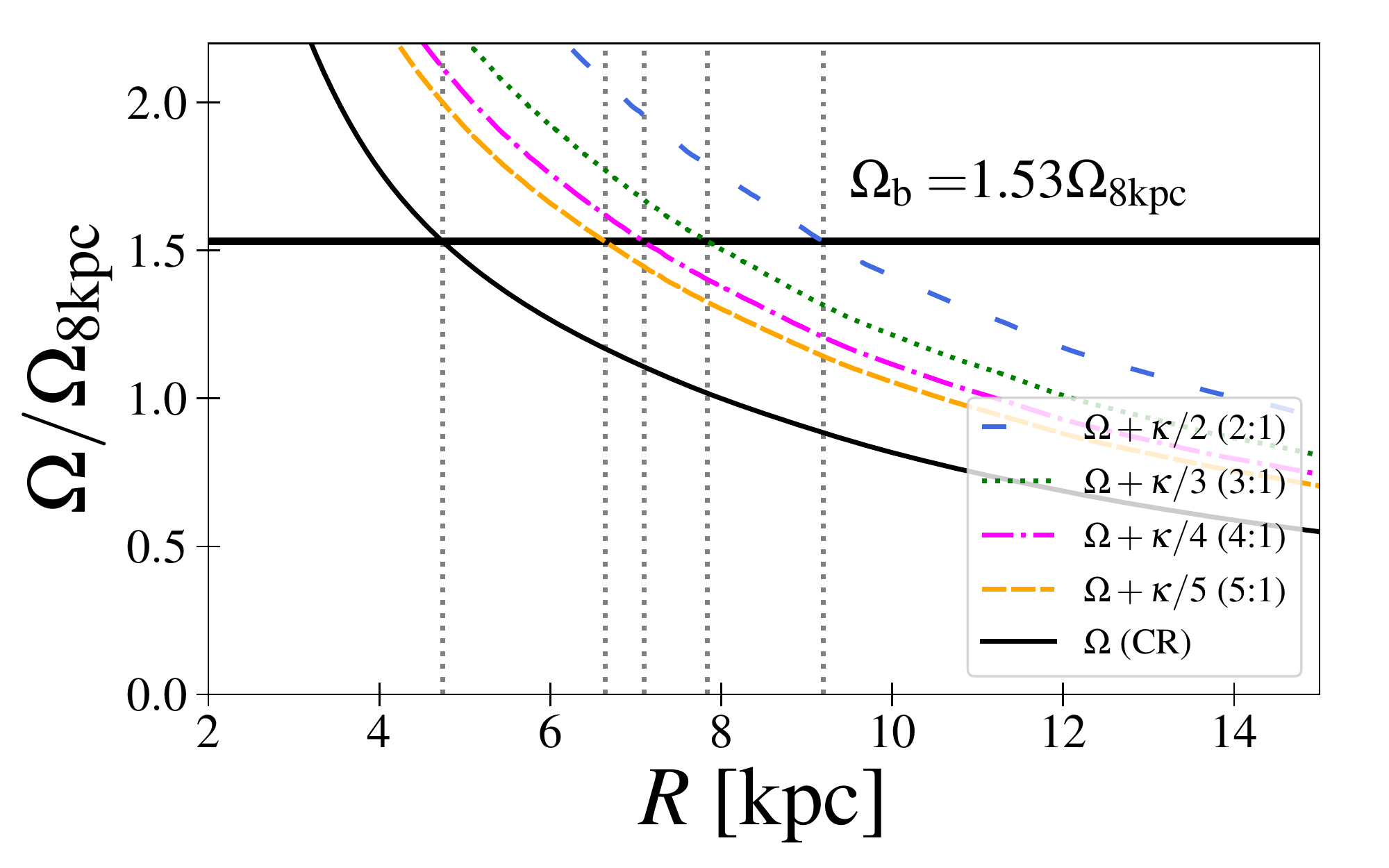}
	\caption{The orbital frequencies as functions of the galactic radius, $R$. $\Omega$ was determined from the velocity curve and $\Omega+\kappa/m$ ($m=$2, 3, 4, and 5) as expected within the epicycle approximation are plotted. The vertical axis is normalized by $\Omega_{8\mathrm{kpc}}$. The horizontal line indicates the bar's pattern speed, and the vertical lines represent the guiding radii for the respective resonant orbits.}
	\label{fig:R_Omega}
\end{figure}

The guiding-center radius ($R_{\rm g}$) for each OLR for our simulated model can be determined as we know the structure of the simulated galactic disk. 
The radial frequency under the epicycle approximation is given by \citep{BinneyTremaine2008}:
\begin{equation}
	\kappa ^2 (R_\mathrm{g}) = 
	{\left( 
			R \frac{\diff \Omega^2}{\diff R} + 4\Omega^2
	\right)}_{R_\mathrm{g}}.
	\label{eq:radial_frequency}
\end{equation}
Here, we calculate $\kappa$ and $\Omega$ at each radius from the particle data of the $N$-body simulation. We also determine the circular frequency, $\Omega$, by averaging the radial accelerations of particles in an annulus with a width of 50 pc at each radius. In Fig.~\ref{fig:R_Omega}, we show $\Omega + \kappa/m$ ($m=$2, 3, 4, and 5) and $\Omega$ as functions of $R$. 
The horizontal line in Fig.~\ref{fig:R_Omega} indicates the pattern speed of the bar. The radii at which the line intersects with the frequency curves correspond to the guiding-center radii for the 2:1, 3:1, 4:1, 5:1, and the corotation resonances. In our model of the Milky way, the Sun, at a distance of $\sim 8$\,kpc from the Galactic center, is close to the 3:1 OLR. It is then not surprising that this resonance is dominant in the local stellar population.

\subsection{The Milky Way's bar pattern speed}
\label{Ssect:Bar}
\begin{figure*}
	\centering
	\includegraphics[width=\columnwidth]{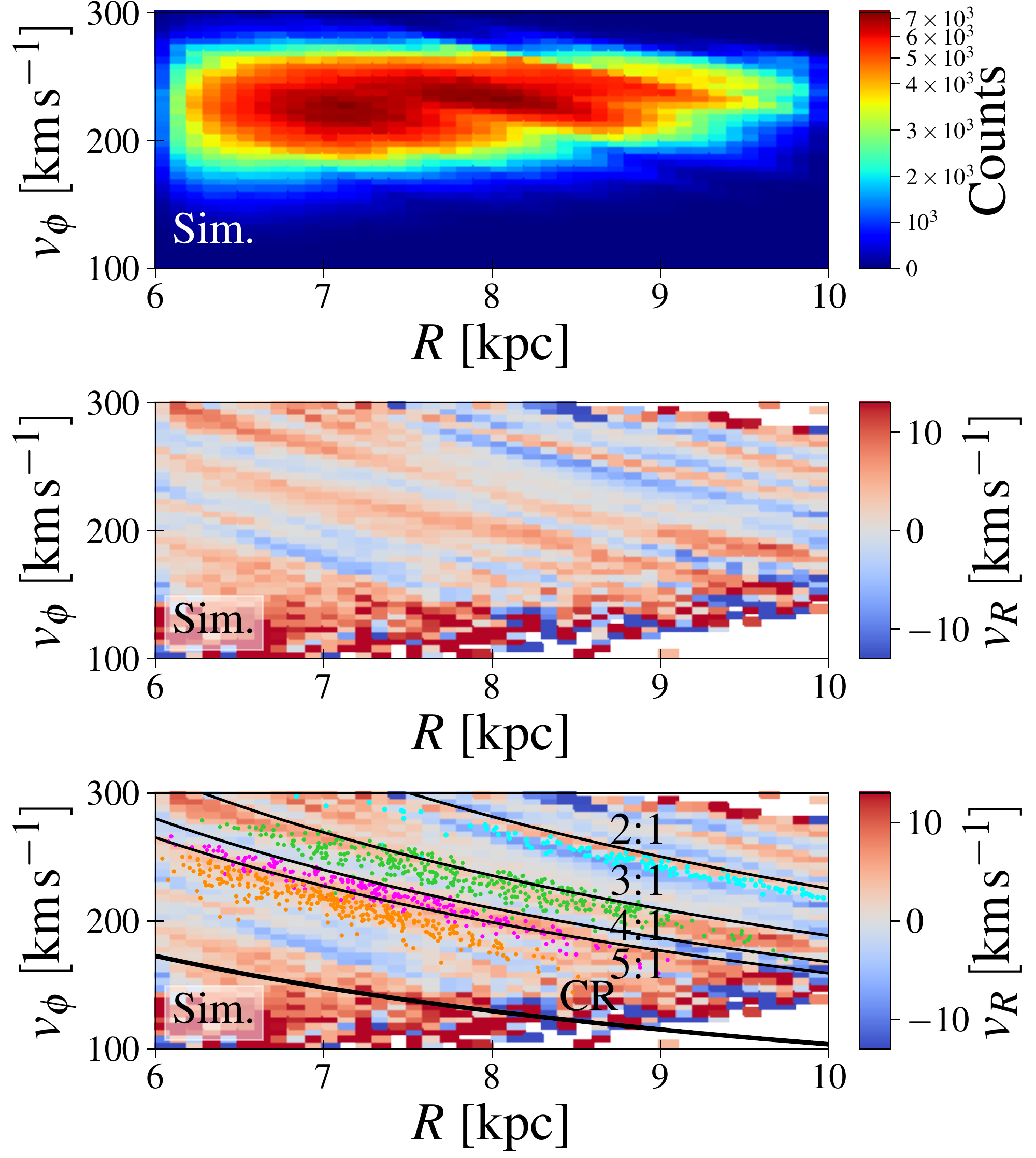}
	\includegraphics[width=\columnwidth]{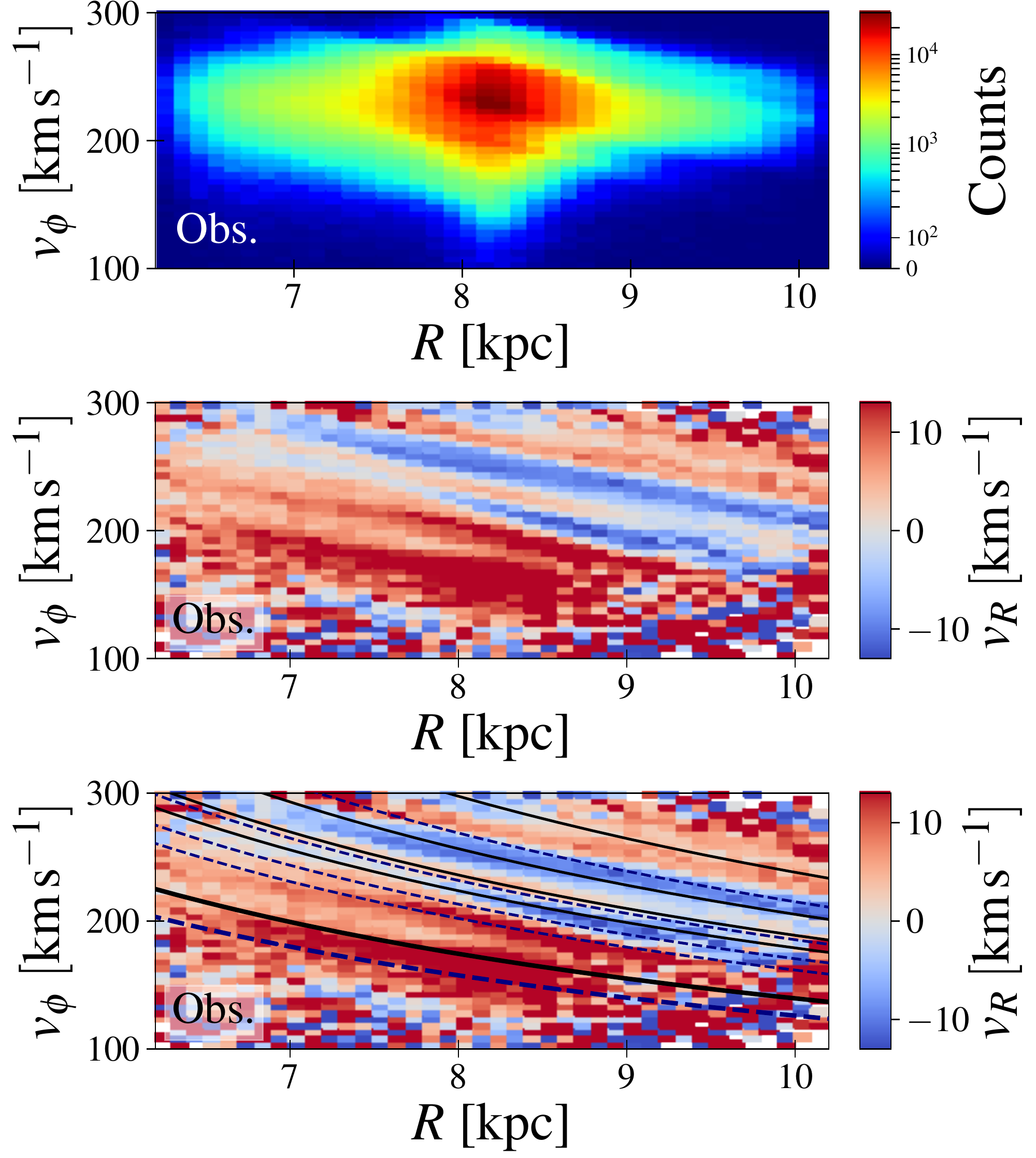}
	\caption{{\it Left:\/} Phase-space structures in the $R$ versus $v_{\phi}$ space from the MW simulation. {\it Right:\/} Counterparts from {\it Gaia} DR2. {\it Top:\/} Colors indicate the number counts in each bin. {\it Middle:\/} Colors indicate the mean $v_R$ in each bin. {\it Bottom:\/} Same as middle panels, but with curves of constant angular momentum overplotted. From the top to bottom curves, they correspond to the angular momenta of the circular orbits at a resonant radius of the 2:1, 3:1, 4:1, 5:1 OLR, and CR. Solid and dashed curves in the bottom right panel correspond to a bar's pattern speed of $\OmegaBar = 1.4 \Omega_0 \simeq 40 \; \mathrm{km\; s^{-1} \; kpc^{-1}}$ or $\OmegaBar = 1.55 \Omega_0 \simeq 45 \; \mathrm{km\; s^{-1} \; kpc^{-1}}$, respectively. Dots in the bottom left panel indicate positions of particles trapped in each of the resonances. The colors are the same as those in Fig.~\ref{fig:uv_resonance}.}
	\label{fig:R_Vphi}
\end{figure*}

When projecting the {\it Gaia} DR2 data into the $R$-$v_{\phi}$ plane, a series of ridges appears over a wide range of $R$ \citep{Kawata+2018,Antoja+2018,Ramos+2018}.
Although the origin of these structures is still in debate, previous studies \citep[e.g.,][]{Monari+2017c, Monari+2019a, Fragkoudi+2019, Barros+2020} have discussed that these ridges may originate from resonances, while other studies have suggested perturbations by a satellite galaxy such as the Sagittarius dwarf \citep{Khanna+2019,Laporte+2019} or by winding transient spiral structure \citep{Hunt+2018,Hunt+2019}.
In this section, we study the relation between the observed ridges in \emph{Gaia} DR2 and resonances in our simulated MW model: by projecting our simulated and classified stars into the $R$-$v_{\phi}$ plane, we can directly compare to the same projection of the {\it Gaia} DR2 data. If the stars classified by resonance type align with the observed ridges, then this provides compelling evidence that these ridges are indeed caused by resonantly trapped stars.

In Fig.~\ref{fig:R_Vphi}, we show the distributions of stars in the $R$-$v_{\phi}$ plane as it arises in our simulated MW model (left column) and in the {\it Gaia} DR2 (right column). 
The top panels show 2D-histograms of particles within 2~kpc from the Sun [$(R,\phi) = (8\, \mathrm{kpc}, 20^{\circ})$, in our model]. This panel reveals a selection effect caused by {\it Gaia}'s measurement uncertainties: due to us having selected stars with relative parallax error below 10\% the right-hand panel displays systematically fewer stars in approximately concentric rings around the Sun's position at $8.2\, \mathrm{kpc}.$ Visually, this causes the smaller extend of the red area in the right-hand panel.
We expected, however, that our analysis is robust with respect to this selection effect: it implies that fewer stars make it into our DR2 projection than into the projection of our simulation, but those stars that are selected are not systematically biased.

The middle panels show the mean $v_R$ values in the respective bins in the top panels.
In order to make this figure, we selected stars from the {\it Gaia} DR2 catalogue whose relative errors in parallaxes are less than 10\%, and whose errors in radial velocities are less than $5 \; \mathrm{km \; s^{-1}}$. Their distances from the Galactic mid-plane are less than 0.2~kpc, and their distances from the Sun are less than 2~kpc.
The total number of stars in this sample is 2,289,755.
We assume that the distance of the Sun from the Galactic center is $R_0 = 8.2$~kpc, and that the distance of the Sun from the Galactic mid-plane is about $z_0 = 25$~pc, 
whereby the velocity of the Sun with respect to the Local Standard of Rest (LSR) is about $(U_{\sun},V_{\sun}) = (10, 11) \, \mathrm{km \, s^{-1}}$, and the circular velocity at $R=R_0$ of $\Theta_0 = 238 \; \mathrm{km \; s^{-1}}$ \citep{Bland-HawthornGerhard2016}.

In the middle right panel in Fig.~\ref{fig:R_Vphi}, we can identify several streams, or ridges, in the \emph{Gaia} data. 
For convenience, we refer to ridges with $v_R>0$  as `red' and $v_R<0$ as `blue'. From top to bottom, red ridges alternate with blue ridges.
Amongst these, two red ridges of $v_R>0$ are located at $v_{\phi} \sim 200 \; \mathrm{km\; s^{-1}}$ and $v_{\phi} \sim 180 \; \mathrm{km\; s^{-1}}$ at 8\,kpc,  are associated with the Hercules stream.
In the middle panel our simulation is shown on the left-hand side. It shows ridges similar to those observed in the  \emph{Gaia} data (middle panel on the right-hand side). In our previous discussion, we argued that the Hercules-like stream in our MW model corresponds to a ridge from $(R, v_{\phi}) \simeq (6\, \mathrm{kpc}, 250 \,\mathrm{km \, s^{-1}})$ to $(R, v_{\phi}) \simeq (9\, \mathrm{kpc}, 180 \, \mathrm{km \, s^{-1}})$. This ridge, eventually branches into two separate structures at around 8\,kpc.

In order to study the relation between the resonances and ridges in our simulation, we overplot the positions of particles in the resonances in the bottom left panel of Fig.~\ref{fig:R_Vphi}. We here randomly sample 5000 particles within 2~kpc from $(R, \phi) = (8\,\mathrm{kpc}, 20^{\circ})$ and extract resonant particles from them as we do in Sec.~3. 
In addition, we plot curves of constant angular momentum ($L_z = R v_{\phi}$) in the figure. 
These correspond to the angular momenta of the circular orbits at the resonance radii determined in Fig.~\ref{fig:R_Omega}.
As discussed in studies such as \citet{Ramos+2018} and \citet{Quillen+2018a}, resonant orbits approximately follow constant $L_z$ curves if their radial oscillations are small.

The overplotting reveals that both the resonant particles and the constant $L_z$ curves follow the ridges. While \citet{Monari+2019a} showed that the constant $L_z$ curves are blue ridges ($v_{R}<0$) for 2:1, 3:1, and 4:1 resonances according to their Figure~6, our $L_z$ curves fall just between red ($v_R>0$) and blue ridges. In addition, we find that the resonant particles are always distributed below the constant $L_z$ curves in $R$-$v_{\phi}$ map. 
This arises due to their radial oscillations not being negligible.
Orbits in a resonance follow a line in $L_z$ vs. $J_R$ space where $J_R$ is a radial action (see Fig.~4 in \citealt{Binney2018}).
As the resonant lines have negative slopes, if resonant orbits $J_R$ values are large, their $L_z$ values are smaller than those of the circular orbits ($J_R=0)$.

Turning to the \emph{Gaia} data, the ridges seem to trace the constant angular momentum lines also in our Milky Way's data.
The solid curves in the bottom right panel of Fig.~\ref{fig:R_Vphi} represent the angular momenta of the resonances for the bar's pattern speed of $\OmegaBar = 1.4\Omega_0$, which is suggested by  \citet{Sanders+2019} and \citet{Bovy+2019}. We also present the same ones but for $\OmegaBar = 1.55\Omega_0$, which is the pattern speed in our simulation. 
In order to determine the radii of the resonances and corresponding angular momenta, we assume a flat rotation curve of $v_{\mathrm{c}}=238\;\mathrm{km \; s^{-1}}$.

Consider that the top red ridge corresponds to the 2:1 OLR, the full curve (at $\OmegaBar = 1.4\Omega_0$) then is close to the relation between the color and constant $L_z$ curve as shown in our simulation (the curve is located just above the red ridge). However, with $\OmegaBar = 1.4\Omega_0=40$\,km\,s$^{-1}$\,kpc$^{-1}$, there are no OLR around the Hercules stream, but the corotation resonance is located at the lowest value of $v_{\phi}$ of the Hercules stream.
Accordingly, if we assume a higher pattern speed, i.e., $\OmegaBar = 1.55\Omega_0 = 45$\,km\,s$^{-1}$\,kpc$^{-1}$, the 2:1 resonance is located slightly below the top red ridge, but the relation between the resonances and ridges for the 4:1 and 5:1 resonances looks similar to that seen in our simulations. In both pattern speeds, the pronounced red ridge near the bottom which we identify with the third Hercules stream, correspond to the corotation resonance. We find the same phenomena in our simulation, but without the stars in corotation resonance.
The latter is a consequence if the relatively low-resolution of our simulations and the fact that in the \emph{Gaia} data the corotation resonance is closer to the Sun than in our simulations.
This is a consequence of our models not perfectly matching to the real Milky Way galaxy.
Thus, our comparison between the \emph{Gaia} data and our simulation suggest that the pattern speed of the Milky Way's bar is relatively slow, with $\OmegaBar = 1.4$--$1.55\Omega_0$ which corresponds to 40--45\,$\mathrm{km \; s^{-1} \; kpc^{-1}}$)


\section{Summary}
We have analyzed a finely resolved $N$-body simulation of a Milky Way-like galaxy obtained in \citet{Fujii+2019} in order to investigate the origins of the phase-space structures in the Milky-Way galaxy as observed by {\it Gaia}. 
We investigated the distribution of particles in the $v_R$-$v_{\phi}$ plane by iterating over multiple positions in the disk. This revealed  
a Hercules-like stream around $(R,\phi) = (8\, \mathrm{kpc}, 20^{\circ})$ in our simulation, as is also known from the actual solar neighborhood. 
From a spectral analysis of stellar orbits, we found mainly four resonances, namely the 2:1, 3:1, 4:1, and 5:1 resonances around there.
The observed structures in the $v_R$-$v_{\phi}$ plane, can be explained by stars being trapped in the 4:1 and 5:1 outer Lindblad resonances. In our simulations, these resonances give rise to structures similar to those observed in the actual Hercules stream.
Our results therefore favor that the Hercules stream is composed out of stars trapped in the 4:1 and 5:1 OLR and CR, which would also explain its trimodal structure as revealed by \emph{Gaia} DR2. In addition, the stars in the 2:1 and 3:1 OLR match to the `hat' and `horn' structures, respectively.

We further compared the distribution of stars in the $R$-$v_{\phi}$ plane in our simulation with that obtained from the {\it Gaia} data. Particles identified to be in resonance in our simulation follow ridges in the $R$-$v_{\phi}$ plane. Similar ridges have also been found in the \emph{Gaia} data and matching the observed ridges to resonances, our results suggest a relatively low pattern speed of the Milky Way's bar, namely $\OmegaBar = 1.4$--$1.55\Omega_0$, which corresponds to 40--45\,$\mathrm{km \; s^{-1} \; kpc^{-1}}$. This is consistent with recent studies \citep[e.g.][]{Sanders+2019,Bovy+2019}.

In contrast to test particle models using static potentials, $N$-body models have some difficulties when comparing the results with observational data due to the time-dependence of the bar and spiral arms.  \citet{Fujii+2019} showed that $v_R$-$v_{\phi}$ maps obtained from $N$-body simulations change with time. This is a natural consequences of dynamic spiral arms \citep[e.g.][]{1984ApJ...282...61S,Baba+2013}. 
In fact, \citet{Hunt+2018, Hunt+2019} discussed phase mixing by dynamic transient spiral arms, which in combination with bar resonances, creates a velocity-space distribution similar to what is seen in observations \citep[see also][]{DeSimone+2004}. 
Furthermore, the evolution of the bar can also affect the stellar velocity-space distribution. Recently, \citet{Chiba+2019} showed that stars trapped in the resonances of the bar are dragged in the phase space when the bar slows down using secular perturbation theory and test particle simulations \citep[see also][]{Weinberg1994,Halle+2018}. This can also cause the formation of a Hercules-like structure. 
We will discuss the time dependence of the velocity distribution, caused by the dynamical evolution of spiral arms and bar in fully self-consistent $N$-body simulations, in a forthcoming paper.

\section*{Acknowledgements}
We thank the anonymous referee for the useful comments.
We thank Kohei Hattori and Anthony Brown for helpful discussions.
This work was supported by JSPS KAKENHI Grant Nos. 18K03711, 18H01248 and 19H01933, and the Netherlands Research School for Astronomy (NOVA). 
MF is supported by The University of Tokyo Excellent Young Researcher Program.
Simulations are performed using GPU clusters, HA-PACS at the University of Tsukuba, Piz Daint at CSCS, Little Green Machine II (621.016.701) and the ALICE cluster at Leiden University. Initial development has been done using the Titan computer Oak Ridge National Laboratory. This work was supported by a grant from the Swiss National Supercomputing Centre (CSCS) under project ID s548 and s716. This research used resources of the Oak Ridge Leadership Computing Facility at the Oak Ridge National Laboratory, which is supported by the Office of Science of the U.S. Department of Energy under Contract No. DE-AC05-00OR22725 and by the European Union’s Horizon 2020 research and innovation programme under grant agreement No 671564 (COMPAT pro ject).
This work has made use of data from the European Space Agency (ESA) mission {\it Gaia} (\url{https://www.cosmos.esa.int/gaia}), 
processed by the {\it Gaia} Data Processing and Analysis Consortium (DPAC,
\url{https://www.cosmos.esa.int/web/gaia/dpac/consortium}). 
Funding for the DPAC has been provided by national institutions, 
in particular the institutions participating in the {\it Gaia} Multilateral Agreement.

\section*{Data availability}
The data underlying this article will be shared on reasonable request to the corresponding author.



\bibliographystyle{mnras}
\bibliography{ms,hercules_ref}




\bsp	
\label{lastpage}
\end{document}